# Shape-Dependent, Deep-Learning-Assisted Metamaterial Solid Immersion Lens (mSIL) Super-Resolution Imaging

Baidong Wu[1], Fiza Khan[1], Lingya Yu[2], Zengbo Wang[1]

[1]School of Computer Science and Engineering, Bangor University, Bangor LL57 1UT, UK

[2] Research Centre for Intelligent Healthcare, Coventry University, CV1 5FB, Coventry, UK

## Abstract

We present the first systematic comparison of three $TiO_2$ metamaterial solid immersion lens geometries—sub-hemispherical, super-hemispherical, and full-spherical—for label-free super-resolution imaging. Using SEM, we characterised both the cap profiles and the nanoparticle–fluid immersion at the lens–sample interface, revealing that super-hemispherical lenses achieve the deepest immersion and closest contact with sample features. Imaging experiments under wide-field and laser-confocal microscopes show that this enhanced immersion drives superior resolution and contrast. In addition, we introduce a deep-learning approach based on a SinCUT image-translation model to establish a cross-modal mapping between SEM morphology and optical imaging response, enabling virtual optical predictions and providing a first step toward a digital-twin representation of mSIL imaging behaviour. Electromagnetic simulations further confirm a direct correlation between immersion depth and far-field main-lobe intensity. Our findings demonstrate that careful control of lens shape and nanoparticle–fluid penetration together with data-driven modelling is essential to maximise super-resolution performance in $TiO_2$ mSILs.

## 1. Introduction

Conventional optical microscopy has been an essential tool in modern science and industry since its invention by Leeuwenhoek in the 1670s. In recent decades, it has been widely used in life sciences, materials science, and semiconductor manufacturing, among other fields [1–4]. However, the inherent imaging mechanism of such microscopy limits its ability to characterize objects smaller than 200 nm. According to Abbe's diffraction limit, a conventional lens-based microscopy system cannot resolve two objects separated by a distance smaller than approximately half the wavelength of the light source in air [5]. Fluorescence-based super-resolution imaging techniques were developed to overcome this limitation [6]. Although these technologies offer imaging resolutions down to 1 nm, their application is restricted to samples that can be labelled with fluorophores. In addition, fluorescent dyes may be harmful to live samples and can alter the intrinsic properties of biological specimens [7].

To address these challenges, label-free super-resolution imaging techniques such as near-field–enhanced dielectric microsphere-assisted microscopy (MSM) and hyperlens-assisted microscopy have been widely studied [8,9]. However, hyperlens-based microscopy, which relies on complex metal–dielectric composite structures, suffers from significant optical losses and requires sophisticated and costly nanofabrication, making it less practical than MSM [10]. MSM and its related variants, in contrast, offer high-resolution imaging under broadband illumination by optimising material selection, immersion media, and microsphere geometry at relatively low cost. For example, transparent high-index microspheres such as polystyrene (PS) in air, fused silica ($SiO_2$) in water, and barium titanate ($BaTiO_3$) in water have achieved resolutions of 120 nm, 50 nm, and 40 nm, respectively [11,12].

Beyond the use of prefabricated microspheres, transfer-printing techniques have been explored to fabricate hemispherical and super-hemispherical lenses with controllable placement and potential scalability [13,14]. However, the resolution of such lenses is constrained by the relatively low refractive index of typical printing materials. Focused ion beam (FIB) milling offers an alternative route by precisely shaping high–refractive-index materials, but the method is experimentally demanding and unsuitable for large-scale fabrication [15].

A practical and low-cost alternative uses high–refractive-index titanium dioxide ($TiO_2$) nanoparticles to form sub-hemispherical (sub-mSIL) and super-hemispherical (super-mSIL) lenses through nano–solid–fluid assembly (NSFA) [16]. Under white-light illumination, $TiO_2$ nanoparticle-based super-mSILs have demonstrated resolutions down to ~45 nm with high contrast and a wide virtual field of view. Two mechanisms were proposed to explain this behaviour. First, the flat bottom of the super-mSIL enhances evanescent-wave reception and suppresses lateral movement of the microlens, preventing air gaps that would otherwise exceed the evanescent penetration depth [17]. However, when illuminated with a monochromatic source, the theoretical lateral resolution of an ideal homogeneous super-hemisphere should be significantly worse than the experimentally observed performance of $TiO_2$-based super-mSILs, despite their similar geometry. SEM characterisation of the mSIL bottom surface revealed that $TiO_2$ nanoparticles penetrate deeply into the underlying nanostructures, filling nanoscale gaps and locally enhancing near-field coupling. Full-wave three-dimensional simulations have shown that such nanoparticle penetration can illuminate nanostructures more effectively and strengthen focused evanescent fields, although earlier simulations did not explicitly incorporate the penetration process.

Building on the $TiO_2$-based NSFA method, our group recently developed a simpler and more repeatable fabrication protocol for micro-full-spherical (full-mSIL) lenses. These lenses exhibit higher imaging contrast, a wider field of view, and better resolution than prefabricated $BaTiO_3$ microspheres. However, the penetration behaviour of $TiO_2$-based full-mSILs has not yet been investigated, and no comparative study has evaluated the imaging performance of sub-, super-, and full-mSILs fabricated under identical conditions.

In this work, we systematically compare the imaging performance of these three geometries using wide-field and laser-confocal microscopy, supported by SEM characterisation of their bottom surfaces. Wide-field imaging reveals that the super-mSIL provides the best resolution (down to ~60 nm), the highest contrast, and a moderate field of view and magnification. Full-wave simulations further show that nanoparticle penetration enhances the collection of propagating-wave energy in the far field, with the super-mSIL benefiting most from this effect.

In addition to the experimental and numerical investigations, we introduce a deep-learning framework motivated by a practical limitation of mSIL imaging. Because each mSIL becomes immobilised at the location where it forms, it cannot be repositioned to image multiple spatial regions, making large-area mapping challenging. Recent advances in deep-learning-based microscopy and cross-modality image translation [21–23] offer opportunities to overcome this constraint. Here, we employ a contrastive unpaired image-translation model to learn a mapping from SEM morphology to optical response, enabling virtual prediction of mSIL imaging at previously unmeasured positions and establishing the basis of an mSIL digital-twin concept.

Together, these contributions provide both mechanistic insight into how lens geometry affects super-resolution performance and a complementary computational route for extending the effective imaging capability of $TiO_2$ mSILs.

## 2. Materials and Methods

Figure 1 illustrates the bottom-up fabrication process of mSIL lenses using $TiO_2$ nanoparticles with an average diameter of 20 nm and a refractive index of 2.5 (supplied by XuanChengJingRui New Material Co. Ltd., Anhui, China). This work builds upon our previously established bottom-up fabrication method, with the addition of a centrifugation step designed to remove oversized and irregular nanoparticles. Aqueous suspensions of 20 nm anatase-phase $TiO_2$ nanoparticles were centrifuged three times at 10 °C, with lower speeds compared with previous protocol of 17,500 RPM, 20,000 RPM, and 21,000 RPM for 20 minutes, 20 minutes, and 90 minutes, respectively. The precipitates collected after the first and second centrifugation steps, which contained uneven nanoparticles and impurities, were discarded. Following the third centrifugation, the supernatant was carefully removed to avoid diluting the tightly packed nanoparticle precipitate at the bottom of the tube. Subsequently, deionized water—at a mass equal to half that of the nanoparticle precipitate—was added to the tube for dilution, and the mixture was left undisturbed for 3 hours to allow the formation of a $TiO_2$ gel. The $TiO_2$ gel was then loaded into an air-brush and sprayed at 45° onto the surface of a of silicon chip coated with a thin layer of an organic mixture consisting of hexane and tetrachloroethylene with a volume ratio of 1:2. The organic mixture thin layer is indispensable, as the formation of mSIL lenses relies on a phase transition driven by the combined effects of gravity and oil/water interfacial tension [16, 18]. This transition occurs when the dispersed droplets of the nano–solid-fluid are sprayed onto the surface of a silicon chip with precoated organic mixture layer. Therefore, precisely controlling the duration of this transition is crucial for shaping the mSIL lenses: a longer transition time tends to produce lenses with a higher height-to-width ratio, eventually approaching a full-spherical geometry. Organic mixture layers with thicknesses of approximately 300 μm, 600 μm, and 900 μm were used, corresponding to evaporation times of 3, 6, and 9 minutes, respectively, to form sub-mSIL, super-mSIL, and full-mSIL lenses. In addition, the low surface tension of hexane facilitates nanoscale penetration of the mSILs. In this work, an air gun pressure of 1.0 bar was used, which has been shown to produce a moderate number of mSILs with an average diameter of approximately 20 μm [18].

The fabricated mSILs were initially characterized by SEM and the stage was tilted to 80° to determine the geometry (Zeiss EVO 10, with accelerating working voltage of 20 kV). Afterwards, fabricated mSILs were examined through Olympus DSX-1000 wide-field microscope (40X, NA=0.8 at 455 nm white-light peak wavelength) with  and Olympus OLS-5000 laser confocal microscope (100X, NA=0.95 at 405 nm). For full-mSILs lenses, they were embedded in polydimethylsiloxane (PDMS) layer with thickness of approximately 300 μm to reduce strong reflection caused by the refractive index mismatch between the $TiO_2$ full-mSILs lenses and air. This embedding enhances near-field optical coupling and ultimately enables super-resolution imaging [18]. The SEM characterization of the bottom surface of the mSILs was performed by detaching the lenses from the silicon chip using double-sided adhesive tape, followed by coating them with a 10 nm thick gold layer [**Machine name missing**].

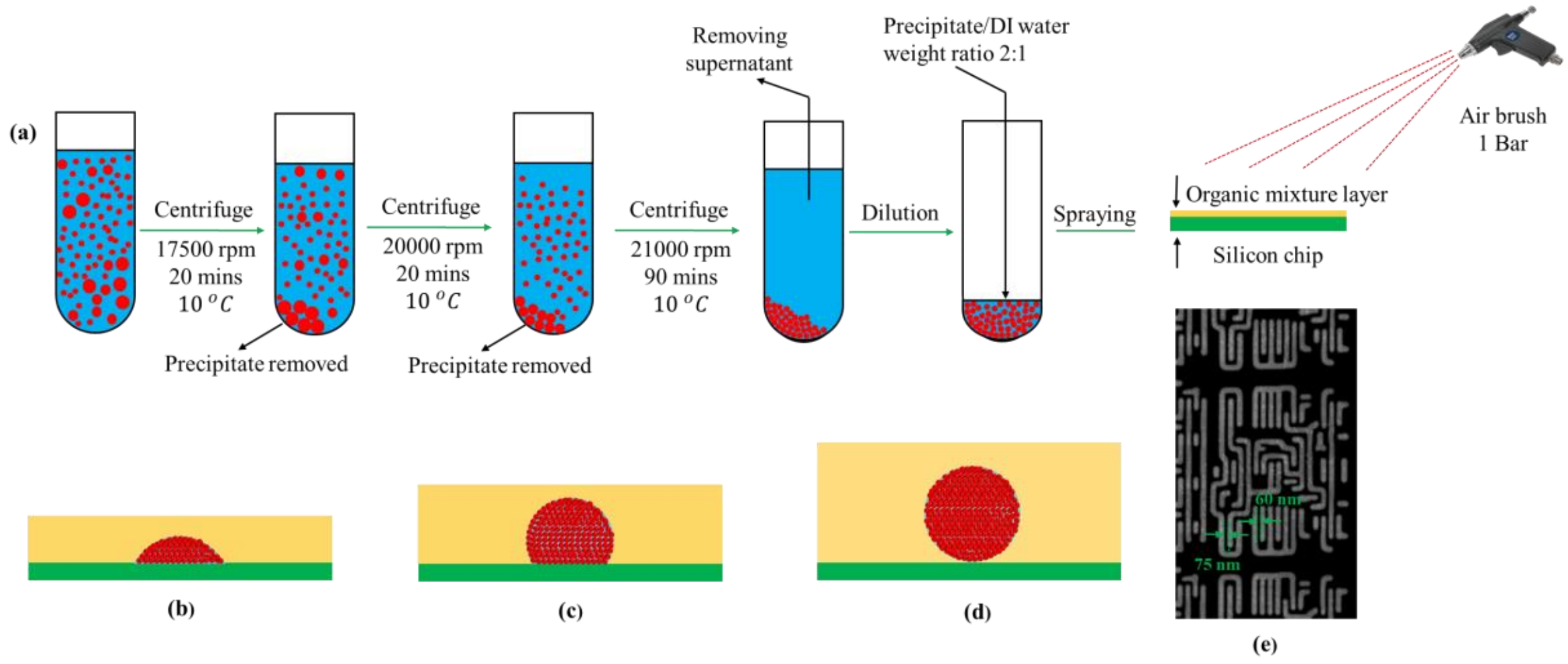

*Figure 1 Schematic illustration of the fabrication procedure for $TiO_2$ nanoparticle-based mSILs (a), and the dependence of mSIL geometry on the thickness of the organic mixture layer, from thin to thick, corresponding to (b) Sub-mSIL, (c) Super-mSIL and (d) Full-mSIL, respectively. (e) Silicon chip with nanoscale surface patterns, onto which the mSILs are sprayed during the fabrication process.*

## Results

- **SEM characterization results**

Oblique side views of the three mSILs (sub-mSIL, super-mSIL, and full-mSIL) are shown in Figure 2 (a), (e), and (i), respectively, with height-to-width ratios of 0.3, 0.8, and 1,0. This variation in aspect ratio demonstrates that the shape of the mSILs is strongly dependent on the phase transition duration, which is controlled by the thickness of the organic mixture layer pre-coated on the silicon chip.

In addition, as shown in Figure 2 f, the bottom surface characterization of the super-mSIL lens confirms that the $TiO_2$ nanoparticles have fully penetrated into the nanoscale grooves, finally forming a corresponding texture on the bottom surface. This effect is consistent with previous experimental observations. Although a similar penetration effect is observed in the bottom surface of the sub-mSIL lens (Fig. 2b), numerous defects are present. These defects are likely caused by the much shorter phase transition duration, which subsequently has a negative impact on imaging quality [16]. For full-mSIL lens in Figure 2j, to the best of our knowledge, no prior characterization to the bottom surface of full-spherical mSILs has been reported. In our previous work, experimental results suggested that $TiO_2$-based full-mSIL lens demonstrated higher imaging resolution compared to the pre-fabricated BTG full-mSIL lens, despite both having similar effective refractive indices. This conclusion was based on near- and far-field optical responses of a $TiO_2$-based full-mSIL lens obtained through full-wave simulations, which did not characterize the bottom surface or include the nanoparticle penetration effect. In this work, Figure 2j clearly reveals a small penetration area, where the silicon chip texture imprinted on the bottom surface. This structural evidence provides another crucial explanation for the better imaging resolution of $TiO_2$-based full-mSIL lens. The explanation of how the penetration effect enhances imaging resolution will be discussed in the discussion section.

- **Wavelength-Dependent Imaging Performance in Wide-field Imaging**

**Figure 2 shows** wide-field imaging results of a semiconductor chip captured using a $TiO_2$-based super-mSIL lens under white-light illumination. The RGB image (a) and combined red-green (RG) channel (b) reveal clear nanoscale structural features within the dashed region, showing strong agreement with the SEM reference image (c). The image was further decomposed into individual grayscale red, green, and blue channels (d–f), corresponding to typical wavelength bands of approximately **620–750 nm (red)**, **495–570 nm (green)**, and **450–495 nm (blue)**. Among them, the red channel (d) provides the highest contrast and clarity, while the blue channel (f) contributes the least. This enhanced performance in the red channel can be attributed to the lower scattering of longer wavelengths, consistent with Rayleigh's law ($\propto 1/\lambda^4$), which enables deeper penetration and less image degradation from surface nanostructures. These findings highlight the broadband imaging capability of the super-mSIL lens and the importance of wavelength-dependent analysis for optimizing image contrast.

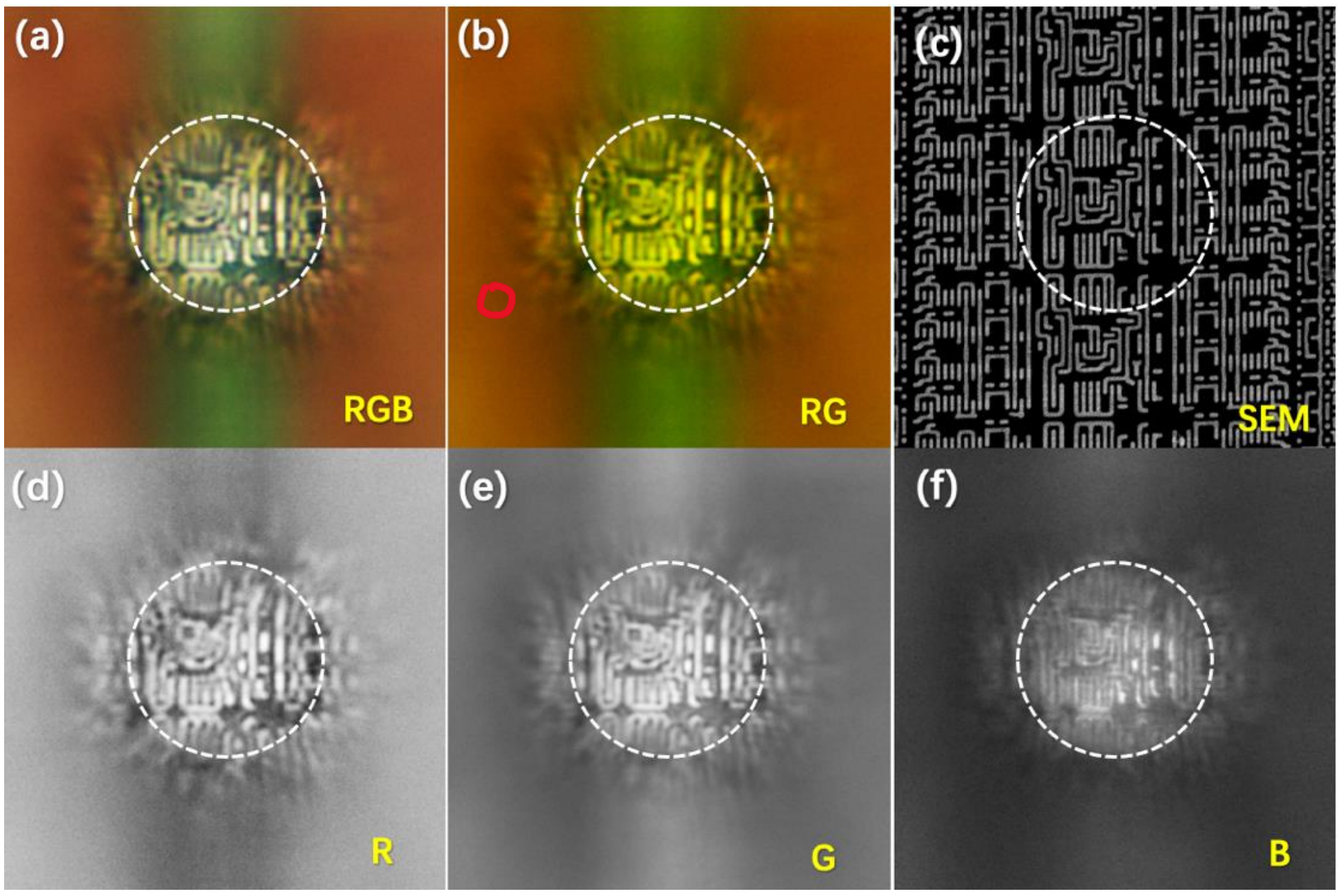

*Figure 2 Wide-field imaging of a semiconductor chip using a $TiO_2$-based super-mSIL lens under white light: (a) RGB image; (b) combined red-green channels; (c) SEM reference; (d–f) individual red, green, and blue channels. Dashed circles indicate regions used for resolution comparison.*

- **Shape effect of mSIL on super-resolution performance**

The silicon chip was imaged through the three mSILs using a wide-field microscope in bright-field mode, as shown in Figure 2c, g, and k, and followed by laser confocal-microscopy in figure 2d, h, and l. The sub-mSIL lens enables the broadest field of view, approximately 11.07

μm$^2$, and followed by super-mSIL and full-mSIL lenses with measured areas of 7.16 μm$^2$ and 4.75 μm$^2$, respectively.

In terms of magnification, the full-mSIL lens provided the highest magnification (5.1X), and decreased to 4.6X and 2.6X for the super-mSIL and sub-mSIL lenses, respectively. The observed decrease in magnification with a reduced height-to-width ratio is consistent with previous research.

Despite not having the highest magnification factor, the super-mSIL lens exhibited the highest imaging resolution and contrast under wide-field microscopy, clearly resolving a 60 nm-wide gap on the silicon chip that is not observable with the full-mSIL or sub-mSIL lenses. Under laser confocal microscopy, this 60 nm gap was subsequently resolved by both super-mSIL and full-mSIL lenses as the single-wavelength laser illumination offers a shorter wavelength and tighter focus, which further improves the system's lateral resolution compared to wide-field microscopy.

The sub-mSIL lens, however, is unable to resolve the 60 nm gap using either of the two microscopy techniques. Under wide-field microscopy, it reveals a clear outline of the chip pattern with secondly high contrast but lacks super-resolution capability. In contrast, under confocal microscopy, it demonstrates modest degree of super-resolution and is able to faintly resolve nanostructures as small as 75 nm, as highlighted in Figure 2d. These results suggest that while the sub-mSIL lens may possible support higher super-resolution imaging quality under confocal microscopy, but it is highly dependent on the quality of penetration effect of bottom surface.

The imaging quality comparison of mSILs fabricated by two-step centrifuge process is shown in Figure S1. At 10°C, the aqueous suspensions of $TiO_2$ nanoparticles was centrifuged at 17,500 RPM and 21,000 RPM for 20 minutes and 90 minutes and all other fabrication processes were kept unchanged. The oblique side-view SEM images show that all three geometric lens types can still be formed by adjusting the phase transition duration. Although confocal microscopy images indicate that mSH and super-mSIL lenses still demonstrate super-resolution capability, neither can resolve 60 nm gap under wide-field microscopy. The sub-mSIL lens can no longer provide a clear outline of nanopattern and does not exhibit super-resolution capability under confocal microscopy. As a result, an additional centrifuge step is necessary to remove medium-sized $TiO_2$ nanoparticles when a lower centrifuge speed is applied.

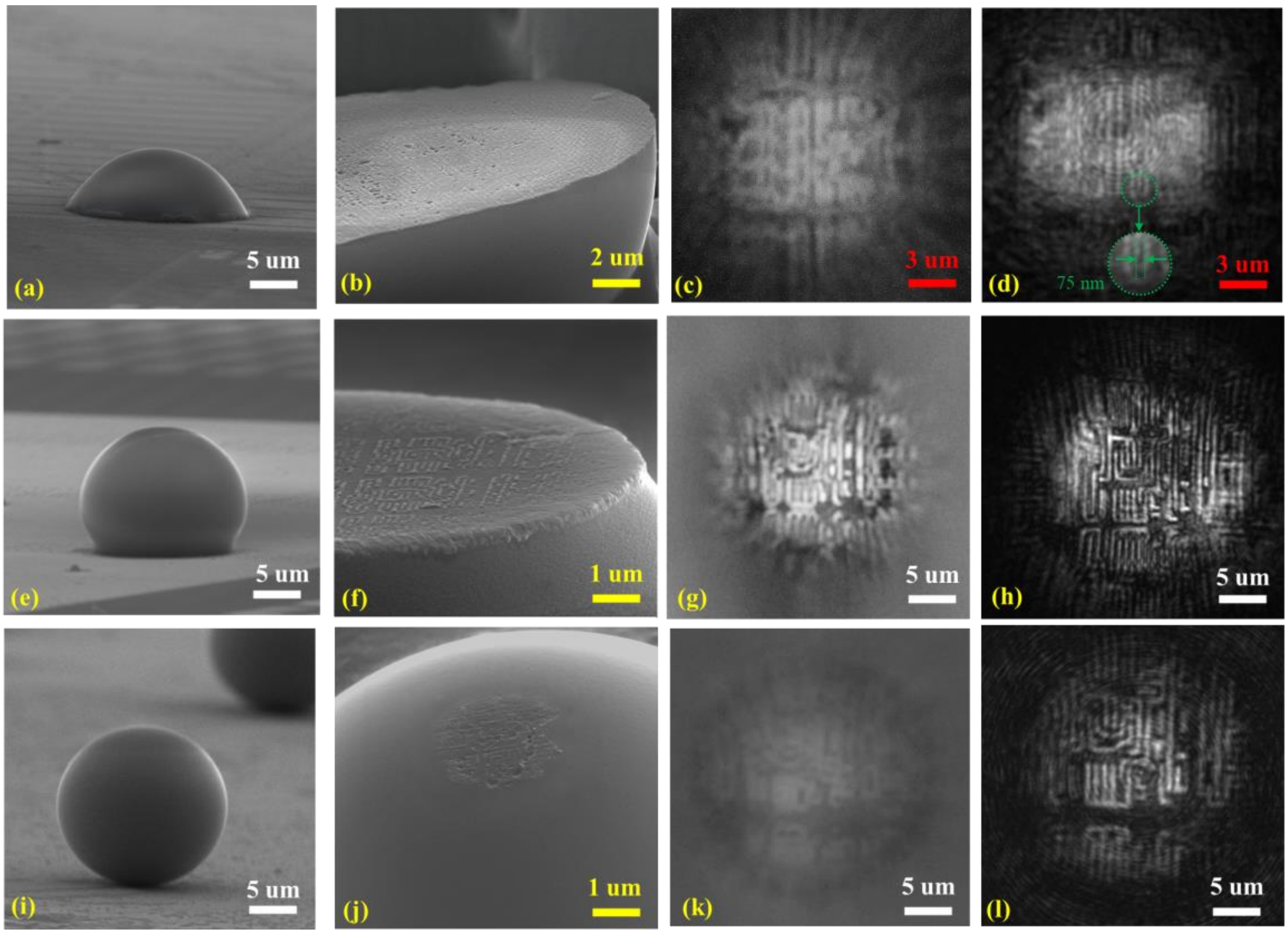

*Figure 3 The SEM images of (a) sub-mSIL, (e) super-mSIL and (i) full-mSIL lenses and their bottom surface characterization (b), (f) and (j), respectively. Wide-field microscopy images, (c), (j) and (k), and laser-confocal microcopy images, (d), (h) and (l) focused on pattern of silicon chip.*

- **Deep Learning–Enabled SEM-Optical Image Translation**

A practical limitation of nanoparticle-based mSIL imaging is that each lens is immobilised at the position where it forms during the nano–solid–fluid spreading process. Unlike prefabricated microspheres, the mSIL cannot be conveniently repositioned to probe different regions of the sample without specialised micromanipulation. Although mechanical or microrobotic actuation could theoretically move the lens, such procedures introduce experimental complexity, risk damaging the nanoparticle penetration interface, and are unsuitable for routine or large-area imaging. In contrast, SEM provides rapid and high-resolution morphological information over the entire chip surface. If a reliable mapping can be established between the SEM morphology (domain A) and the corresponding optical appearance through the mSIL (domain B), then optical responses at arbitrary, unmeasured locations could be reconstructed computationally. In effect, the trained model serves as a digital twin of the physical mSIL, providing a virtual counterpart that predicts how the same lens would image different regions of the substrate without any physical movement. This motivates the use of deep learning–based image translation to generate virtual optical images directly from SEM input, effectively enabling large-area imaging without physically moving the lens.

To achieve this, we employ a Single-Image Contrastive Unpaired Translation (SinCUT) model [20] to learn a cross-modal mapping between SEM and wide-field optical images. SEM and optical images of the same region were first scaled and manually aligned so that the structural features within the view window area were approximately matched. These nearly paired examples were used to train the SinCUT model, which maximises patch-wise mutual information between input and output, ensuring that nanoscale structural details from the SEM are preserved while their appearance is transformed into the corresponding optical domain. Full architecture and training details are provided in Supplementary Information (Section S3).

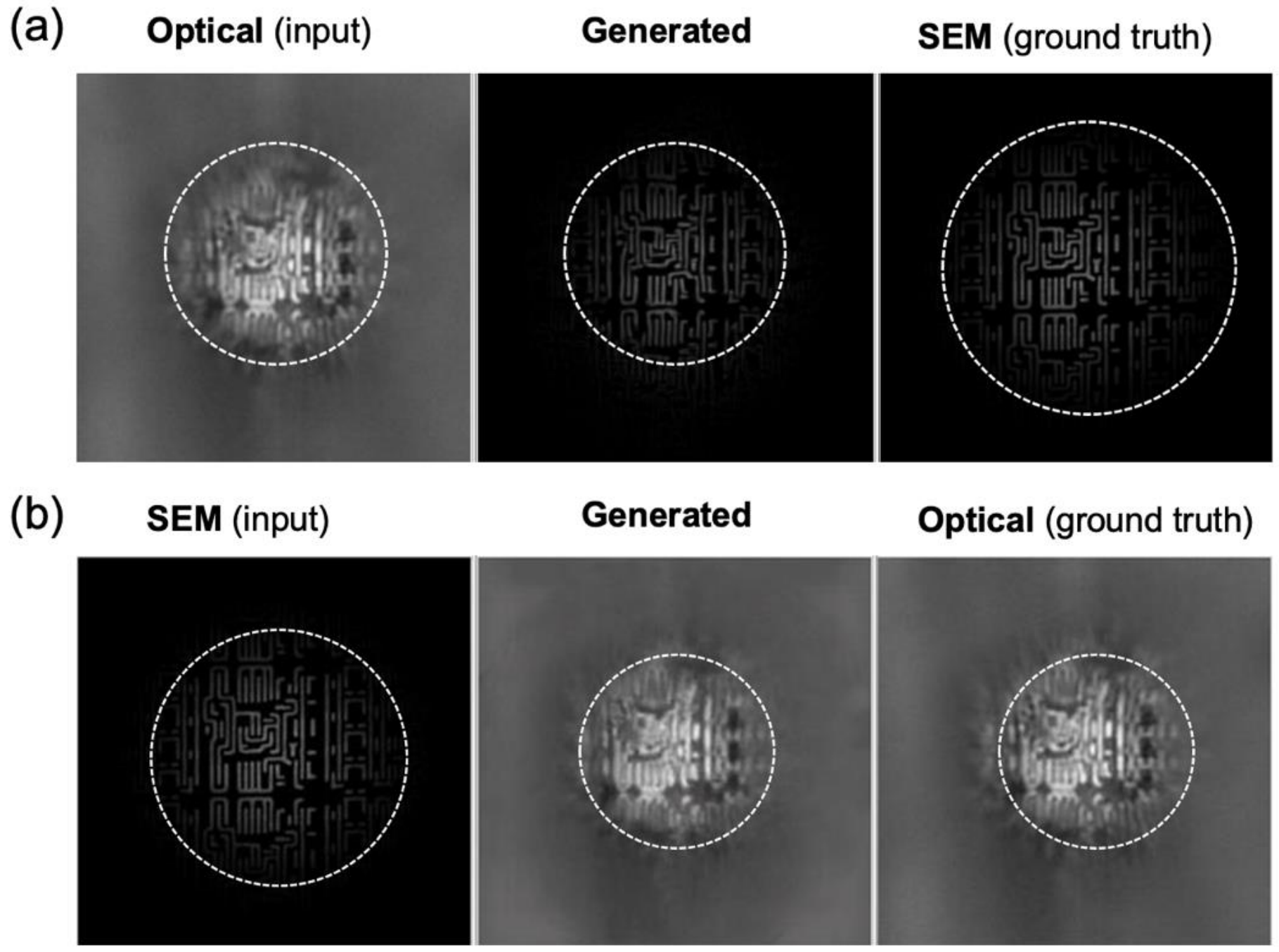

Figure 4 - SinCUT bidirectional image translation. (a) Optical input, generated SEM-like output, and SEM ground truth. (b) SEM input generated optical output, and optical ground truth.

Figure 4 demonstrates the performance of the SinCUT model in translating between the SEM and optical imaging domains. In Fig. 4(a), an optical image is used as input and the model generates an SEM-like output, which closely reproduces the structural geometry and intensity patterns of the true SEM target. Conversely, Fig. 4(b) shows the reverse translation, where SEM morphology is used to predict the corresponding optical appearance through the mSIL. The generated optical image accurately captures the overall contrast distribution, central-zone intensity enhancement, and characteristic distortion features observed in the real optical measurement. Across both translation directions, the model preserves fine nanoscale structures within the dashed region and maintains the geometry-dependent contrast signatures associated with each mSIL type. These results confirm that SinCUT successfully learns a bidirectional mapping between SEM morphology and optical imaging response, enabling faithful cross-modal prediction and supporting the digital-twin interpretation of the mSIL imaging system.

The viewing-window behaviour in Fig. 4 shows an interesting asymmetry: in the Optical→SEM direction (Fig. 4a), the generated SEM-like output largely preserves the view-window size of the optical input, whereas in the SEM→Optical direction (Fig. 4b), the generated optical-like image does not follow the larger SEM window but instead converges toward the characteristic optical window observed in the ground truth. This behaviour reflects a well-known property of contrastive unpaired translation models such as SinCUT. Because patch-wise contrastive learning constrains local structural similarity rather than enforcing global geometric consistency, the model tends to preserve input geometry only when translating **into** a domain whose global appearance is more weakly structured (as in the SEM domain). In contrast, when translating **into** the optical domain, the model adapts the output geometry toward the dominant statistical features of that domain—here, the stable and well-defined bright viewing window typical of mSIL optical images. Despite these geometric adjustments, the generated images in both directions closely reproduce the nanoscale patterns, contrast relationships and structural organisation within the dashed region, demonstrating that the model accurately captures the cross-modal mapping relevant for interpreting mSIL imaging behaviour.

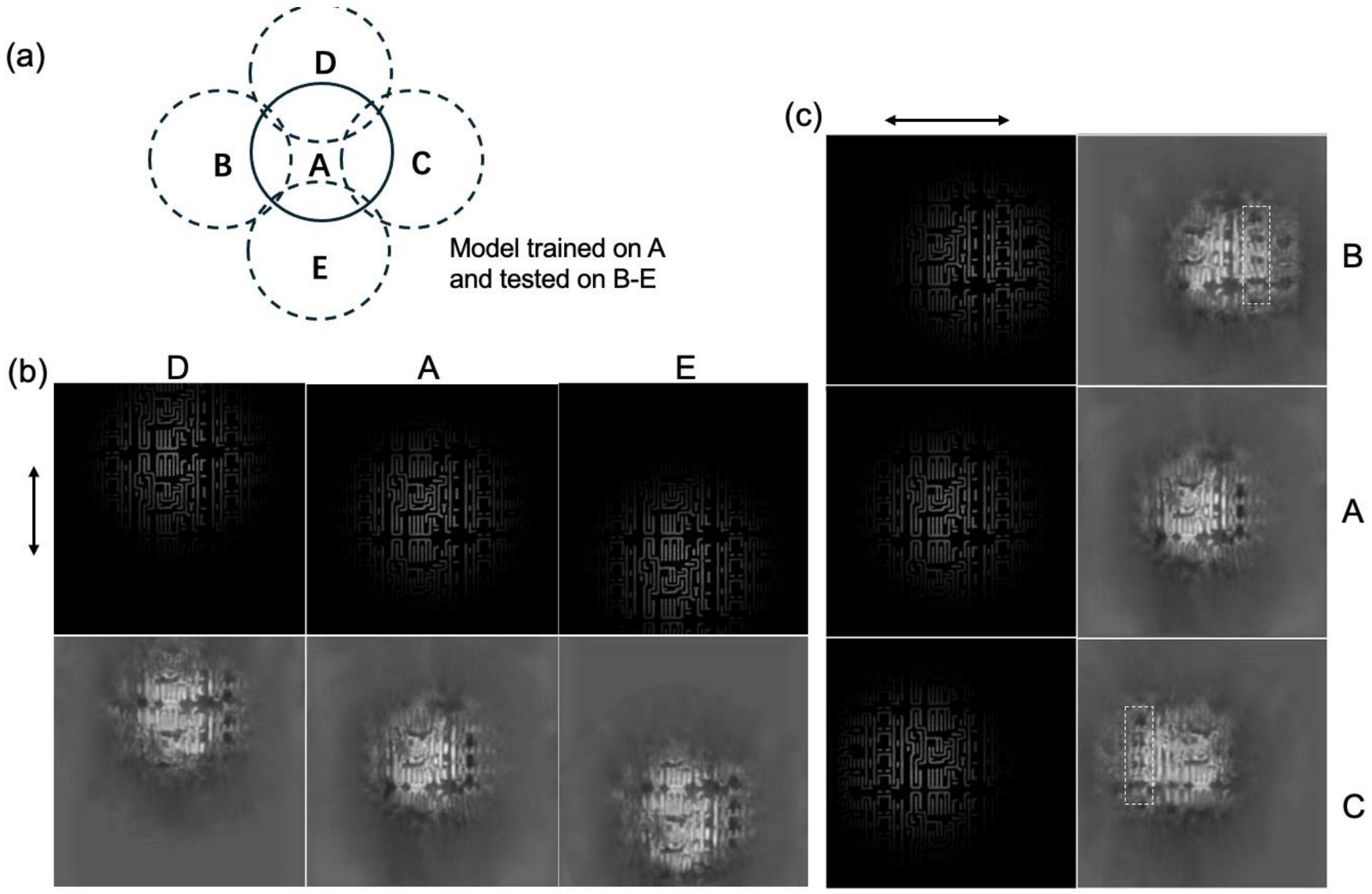

Figure 5- Generalisation of the SinCUT model to unseen locations. (a) Schematic showing the training region (A) and neighbouring test regions (B–E). (b) Predicted optical images for SEM inputs at D–A–E (vertical direction). (c) Predicted optical images for SEM inputs at B–A–C (horizontal direction).

Figure 5 evaluates how the SinCUT model performs when the sample is shifted to neighbouring positions around the central training region A. Panel (a) illustrates the spatial relationship of the five locations: A is the only region used for training, while B, C, D and E are unseen test regions. Panels (b) and (c) show the SEM inputs and the corresponding predicted optical outputs at these locations. When the sample is moved vertically to regions D and E, the reconstructed optical images closely match the expected structure and contrast. This strong generalisation arises because most nanoscale patterns in D and E also appear within region A, allowing the model to reproduce them reliably. For the horizontal positions B and C, the model likewise produces optical predictions that are globally consistent with the expected appearance: the overall brightness distribution, viewing-window shape and major structural motifs are well captured. However, both B and C show a similar local issue: a small portion of the pattern is reconstructed less faithfully (highlighted by the dashed box). This occurs because these specific structural features do not appear in the training region A; with only a single SEM–optical pair used for training, the SinCUT model has no prior exposure to these unseen geometries. Consequently, the model performs well on structures that fall within the statistical distribution of A but struggles with features outside it. Importantly, this limitation does not undermine the value of the approach: the results demonstrate that accurate cross-modal translation is achievable even with minimal training data, establish the feasibility of an mSIL digital-twin framework, and identify pattern diversity—not model capability—as the key factor governing generalisation. Future work will therefore focus on incorporating multiple training regions or applying structural-augmentation strategies to broaden pattern coverage and achieve more complete reconstruction across all unseen locations.

- **Mechanism for shape-dependent imaging performance**

To investigate the imaging mechanism of these mSILs, a full-wave 3D simulation (CST STUDIO, version 2023) was conducted to demonstrate the near- and far-field optical responses. Our previous research suggested that the effective refractive index of the $TiO_2$ nanoparticles close-packed is 1.92 and with a filling factor of 61.2% [18]. Each mSIL is composed of millions of $TiO_2$ nanoparticles. To ensure the computational load remained manageable, the majority of model was simplified as a homogenous material (n =1.92), while the region contacting with nano-pattern was modeled as a $TiO_2$ nanoparticles closely-packed region. In addition, the geometries of those 3D mSILs were simplified by 2D cylinders. As illustrated in Figure 6 a, d, and g, three variants of mSILs with height-width ratio of 0.3, 0.8 and 1.0 were constructed in CST STUDIO with a diameter of 20 μm, incorporating both non-penetration (left) and penetration (right) scenarios. The PDMS used as host medium for full-mSIL lenses is represented in green, and it fills up those empty nanoscale gaps when the penetration effect is not considered. To simulate the experimentally observed partial penetration in full-mSIL lenses, only 3 nanoscale gaps are filled with $TiO_2$ nanoparticles, while remaining gaps are filled with PDMS. For super-mSIL and sub-mSIL geometries, all nano-scale gaps are either filled with air or $TiO_2$ nanoparticles while all other parameters are kept unchanged. The light source is simplified to an electric dipole that is perpendicular to the plane and positioned at the center of the middle gap. This configuration enables us to analyze how evanescent waves generated in the near-field are finally converted to propagating wave in the far-field. The differences in the optical responses between two penetration scenarios, as well as among three variants of mSILs are crucial for explaining their distinct imaging capabilities.

The far-field patterns shown in Figure 6b, e and h suggest that the main characters of far-field, such as main lobe direction and main lobe width, are not significantly affected by penetration effect. However, an enhancement in the main lobe magnitude indicating increased energy collection is observed across all three variants after the penetration effect is considered. The highest enhancement (8.8%), indicating improved energy collection, is observed in the super-mSIL lens and followed by 3.8% and 3.0% for full-mSIL and sub-mSIL lenses, respectively. This trend aligns with experimental super-imaging observations, where the super-mSIL lens exhibited the best super-imaging resolution. Despite the simulated enhancement levels of the full-mSIL and sub-mSIL lenses are similar, in experiments, the poor penetration quality, as shown in Figure 2b, occurring at the bottom of sub-mSIL lens during experiments likely diminishes this enhancement. Notably, the far-field pattern of full-mSIL lens, when the PDMS is introduced as host medium, exhibits multiple sharp peaks rather than single broad main lobe, as observed from super-mSIL and sub-mSIL lenses. This distribution disperses the energy collection of propagating wave in far-field, and may result in lower imaging contrast compared to super-mSIL and sub-mSIL lenses. radiation energy into the far-field

Electric field (E-field) distributions of the three mSIL variants schematically illustrate how radiation energy is coupled into far-field, as shown in Figure 6 c, f, and i, respectively. The magnified image of E-field locating around the penetration area is shown in Figure S2. It clearly shows a photonic nanojets array, which enhances the conversion of the evanescent wave in near-field into the propagating wave in far-field while no significant difference is observed across three mSILs in this aspect.

Overall, we can conclude that the super-resolution imaging capability of $TiO_2$ nanoparticle-based mSILs depends on both their geometries and the penetration quality of the nanoparticles, providing valuable guidance for fabricating and optimizing the performance of other similar mSILs.

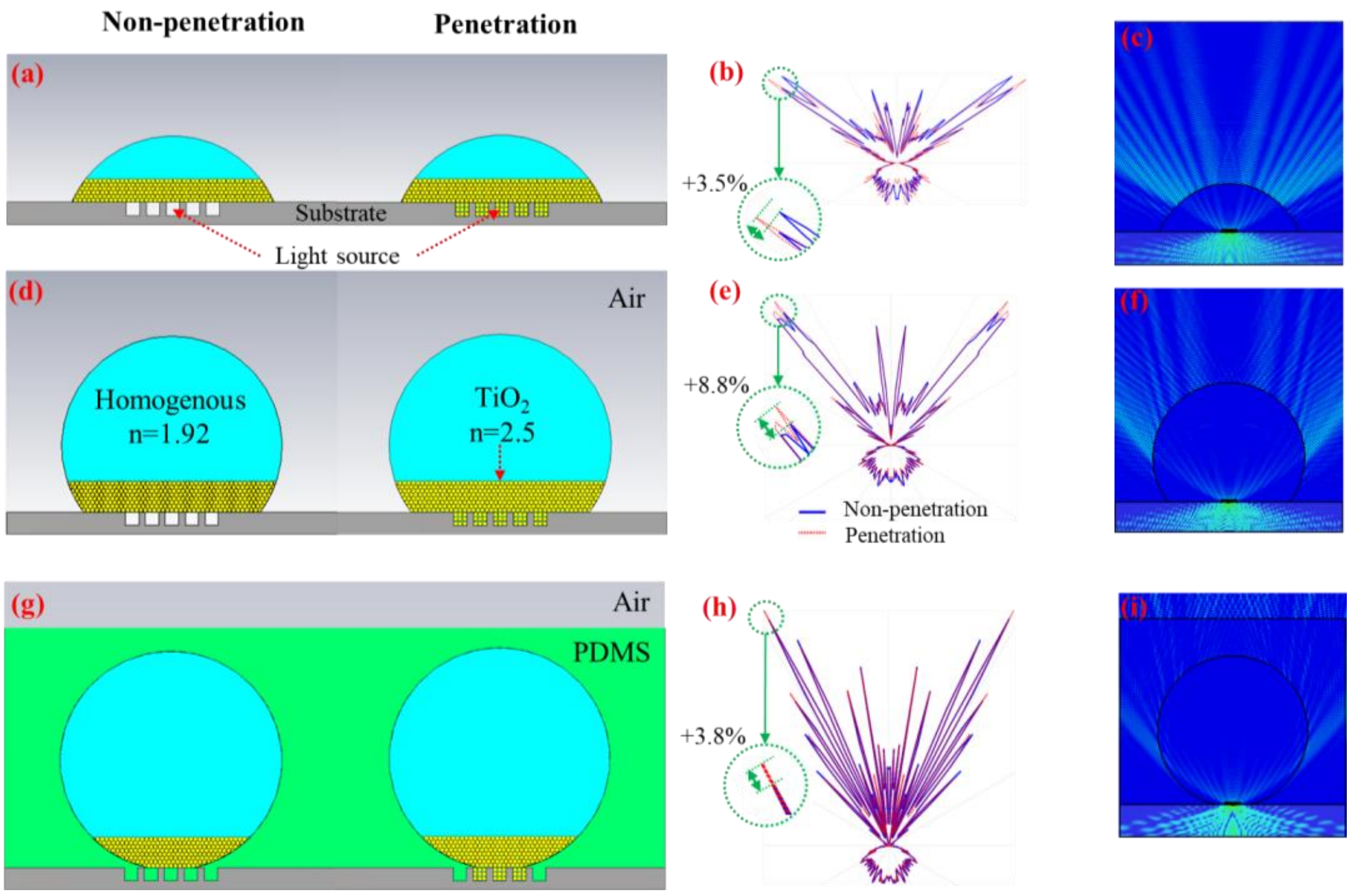

*Figure 6 The schematic illustrations of the simplified models of (a) sub-mSIL, (d) super-mSIL and (g) full-mSIL lenses, respectively, with 20 μm diameters, each incorporating non-penetration (left) and penetration (right) scenarios and their corresponding far-field scattering patterns in (b), (e) and (h). The electric field distributions of three lenses in (c), (f), and (i).*

- Discussion

The combined experimental, computational, and deep learning results provide a unified understanding of how mSIL geometry and nanoparticle penetration jointly determine super-resolution performance. The SEM characterization confirms that super-mSILs consistently achieve the deepest and most uniform penetration into the nanoscale grooves of the substrate, which directly enhances near-field coupling and explains their superior resolution in both wide-field and confocal modalities. The full-wave simulations further show that this penetration effect increases far-field energy collection, particularly for the super-mSIL geometry, in agreement with experimental observations.

The deep learning framework offers an additional perspective on mSIL behaviour. By training a SinCUT model on a single SEM–optical pair, we demonstrate that the SEM→optical mapping can be learned with high fidelity for most nanoscale patterns within the training distribution. The model generalises effectively to neighbouring regions whose structural motifs are also present in the training area, supporting the concept of an mSIL digital twin. At the same time, the reduced reconstruction accuracy for local features that do not appear in the training region highlights the role of pattern diversity, rather than model capacity, as the primary limitation of the current approach. This insight suggests that expanding the training set to multiple spatial locations or enhancing pattern variability through augmentation would yield a more complete virtual imaging framework across larger chip areas.

Taken together, these findings demonstrate that mSIL performance emerges from the interplay between lens shape, penetration quality, and local sample geometry, while also establishing a promising computational direction for digitally extending the lens's effective field of view.

## Conclusion

This work provides the first systematic comparison of $TiO_2$-based sub-mSIL, super-mSIL, and full-mSIL lenses fabricated under identical conditions, revealing clear correlations between lens geometry, nanoparticle penetration, and super-resolution imaging performance. The super-mSIL lens achieves the best experimental resolution and contrast, which is attributed to its large and high-quality penetration area. SEM characterization and full-wave simulations consistently show that the penetration of $TiO_2$ nanoparticles enhances the conversion of evanescent waves into propagating waves, thereby improving far-field energy collection and enabling superior imaging.

In addition to these physical insights, we demonstrate a deep learning–based SEM–to–optical translation framework that acts as a digital twin of the mSIL imaging process. The SinCUT model reconstructs optical responses with high fidelity across most unseen locations, validating the feasibility of computational virtual imaging. The observed limitations at regions containing previously unseen nanoscale geometries highlight the need for expanded training data or structural augmentation, providing a clear path for future improvement.

Overall, this study offers both mechanistic understanding and computational tools for advancing $TiO_2$ mSIL technology, laying the groundwork for scalable, high-contrast, label-free super-resolution imaging systems.

**Supplementary: Models and Training**

# S1. SEM Image Pre-processing

To enable meaningful SEM-to-optical translation using the SinCUT model, the SEM image was pre-processed so that its appearance more closely matches the statistical properties of optical mSIL images while preserving nanoscale geometry. Before filtering, both the SEM and optical images were resized to 256 × 256 pixels and **coarsely aligned**. Coarse alignment is sufficient because precise pixel-level matching is extremely difficult to achieve manually due to differences in magnification, distortion, and illumination. SinCUT operates on local patch statistics rather than fixed spatial correspondences, making coarse alignment adequate.

To emulate the bright-centre and dim-edge illumination pattern typical of mSIL optical images, a radial vignetting mask was applied. For each pixel at (x, y) relative to the image centre ($x_c$,$y_c$), the Euclidean distance

$$d(x,y) = \sqrt{(x-x_c)^2 + (y-y_c)^2}$$

was computed and normalised using

$$r(x,y) = \frac{d(x,y)}{d_{\max}}.$$

A linear attenuation profile

$$M_{\text{vig}}(x,y) = 1 - \alpha r(x,y)$$

was then applied, where α controls the strength of the vignetting.

Optical mSIL images also contain reduced high-frequency detail compared with SEM images. To mimic this behaviour, the vignetted SEM image was smoothed using a Gaussian kernel

$$G(x,y) = \frac{1}{2\pi\sigma^2} \exp\left[-\frac{x^2+y^2}{2\sigma^2}\right],$$

and convolved to produce

$$I_{\text{blur}} = I_{\text{vig}} * G.$$

This filtering preserves the sample geometry while producing an optical-like texture. The resulting pre-processed SEM image was used as the SEM-domain input for SinCUT training. Figure **S1** illustrates the full transformation pipeline.

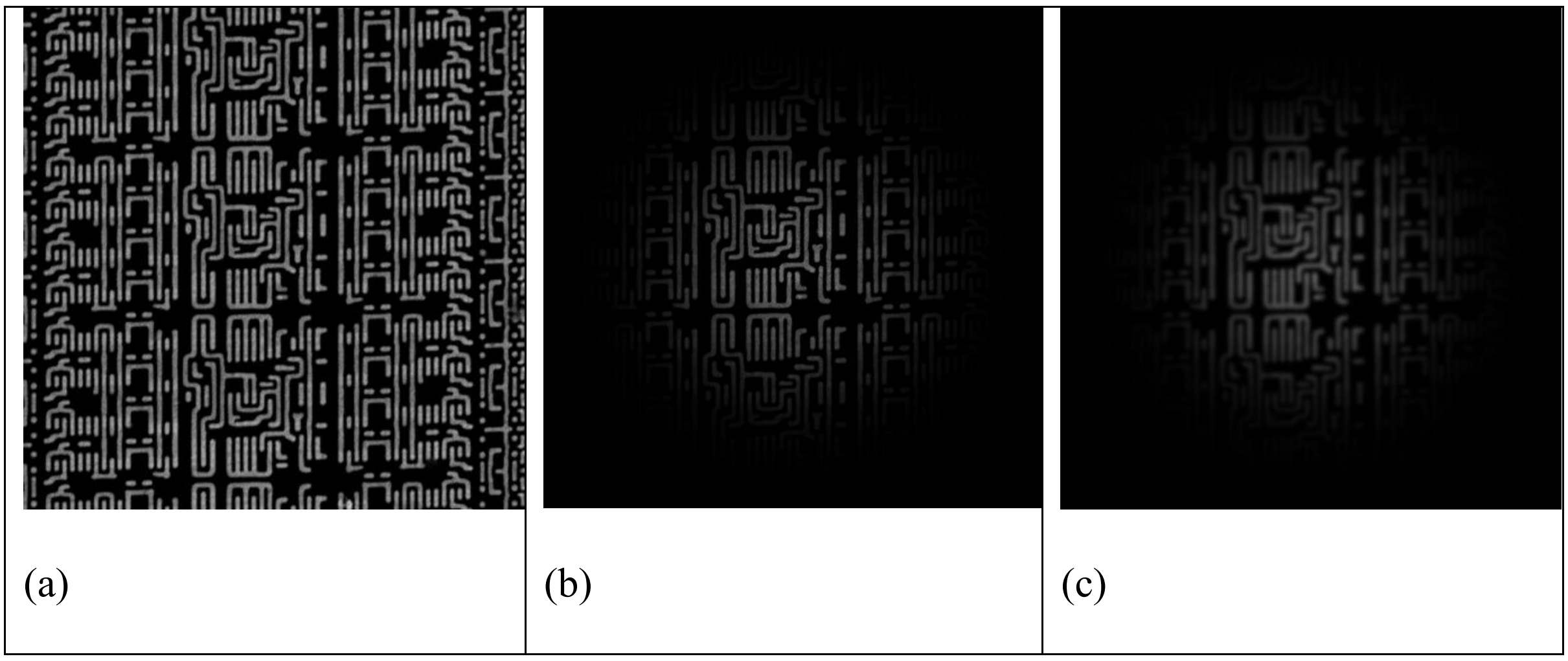

**Figure S1.** SEM pre-processing steps for SinCUT training: (a) Original SEM image, (b) SEM image after applying the vignetting mask, (c) Final SEM image after vignetting and Gaussian blurring, used as the SinCUT input.

# S2. SEM–Optical Translation Using SinCUT

In this work, SinCUT is used solely as a tool to learn a mapping from SEM morphology to the corresponding optical response produced by the $TiO_2$ mSIL. The model is applied without modification, using the standard publicly available implementation. Its role is to generate optical-like predictions for new SEM regions based on the single training pair.

# S3. Model Training Procedure

Training was performed on a single SEM–optical pair resized to 256 × 256 pixels. To manage GPU memory and enhance robustness, the SinCUT model was trained using randomly sampled 128 × 128 crops extracted from the aligned pair during each iteration. This cropping procedure acts as natural data augmentation and ensures that the learned translation does not depend on a fixed spatial offset or magnification. Random scaling was also applied to increase the diversity of training samples.

All training parameters follow the default SinCUT configuration. The model was trained for 16 epochs with a batch size of one and a learning rate of 0.002. All experiments were conducted on an NVIDIA RTX A4000 GPU with 16 GB of memory. This setup yielded stable convergence and produced the cross-modal translation results presented.